# A study of referencing changes in preprint-publication pairs across multiple fields

**Aliakbar Akbaritabar***[1]¶, **Dimity Stephen**[2]¶, **Flaminio Squazzoni**[3]

*[1] Laboratory of Digital and Computational Demography, Max Planck Institute for Demographic Research (MPIDR), Rostock, Germany; ORCID = 0000-0003-3828-1533

[2] German Centre for Higher Education Research and Science Studies (DZHW), Berlin, Germany; ORCID = 0000-0002-7787-6081

[3] Department of Social and Political Sciences, University of Milan, Milan, Italy; ORCID = 0000-0002-6503-6077

* Corresponding author

E-mail: akbaritabar@demogr.mpg.de; akbaritabar@gmail.com (AA)

¶These authors contributed equally to this work

**Abstract**

Manuscripts have a complex development process with multiple influencing factors. Reconstructing this process is difficult without large-scale, comparable data on different versions of manuscripts. Preprints are increasingly available and may provide access to the earliest manuscript versions. Here, we matched 6,024 preprint-publication pairs across multiple fields and examined changes in their reference lists between the manuscript versions as one aspect of manuscripts' development. We also qualitatively analysed the context of references to investigate the potential reasons for changes. We found that 90% of references were unchanged between versions and 8% were newly added. We found that manuscripts in the natural and medical sciences undergo more extensive reframing of the literature while changes in engineering mostly focused on methodological details. Our qualitative analysis suggests that peer review increases the methodological soundness of scientific claims, improves the communication of findings, and ensures appropriate credit for previous research.

1. **Introduction**

Manuscripts have a complex development process with different stages, including peer review at journals. Peer review helps authors to improve the quality of their manuscripts using feedback from previously unrelated experts, who are expected to screen studies for academic rigour, identify their shortcomings and reject those studies that are irretrievably flawed (Atjonen, 2019; De Vries et al., 2009; Dondio et al., 2019). Through the peer review process, journal editors and peer reviewers play key and distinct roles in determining the content of manuscripts (Hojat et al., 2003; Siler et al., 2015). Peer reviewers evaluate the soundness and novelty of a study from their position as topical experts and suggest improvements to the methodology, theoretical framing, and communication of ideas. Journal editors consider the reviewers' input and the study's relevance to their journal to make a decision about publishing (Hirschauer, 2010), ultimately shaping what content and schools of thought constitute the mainstream academic corpus (Hengel, 2017; Hofstra et al., 2020; Teplitskiy et al., 2018; Wu et al., 2017).

To estimate the effects and benefits of peer review, it is vital that we can reconstruct the development of manuscripts throughout the review process. Unfortunately, this requires access to internal journal data that is often impossible to achieve without cooperation from journals. Indeed, barriers and obstacles exist for data sharing between journals and the academic community which would be essential to providing comparative analyses across fields of research (Squazzoni et al., 2020). This has led researchers to estimate the effect of peer review on manuscripts by indirectly gauging the fate of rejected manuscripts eventually published elsewhere (Casnici, Grimaldo, Gilbert, Dondio, et al., 2017). However, this requires extensive data screening and cannot help to estimate the effect of reviewers due to the lack of initial versions of manuscripts and long periods between multiple versions submitted to different journals.

Here, recent open science initiatives can help as they have facilitated the establishment of many preprint servers, where an increasing number of authors deposit their manuscripts prior to journal submission. While preprints are used for rapid dissemination of findings and to establish priority or concurrence (Sarabipour et al., 2019), they could also be used to compensate for the lack of data on initial journal submissions: comparing preprints and their published versions in journals after peer review could help us to at least indirectly estimate the effect of peer review on manuscripts.

For instance, Lin et al. (2020) matched all preprints in computer science submitted to *arXiv* from 2008 to 2017 with their versions later published in journals. They found the published versions included "adequate" revisions and more detailed abstracts and introductions. Given the lack of community involvement in prepublication review (Anderson, 2020), tracing preprint-publication pairs and using preprints as proxies of the initial versions of manuscripts submitted to journals can help to assess the potential impact of peer review on manuscript development (Larivière et al., 2014). This type of research will never permit a precise estimate of the effect of peer review in specific journals as manuscripts could have been eventually rejected multiple times from various journals. However, this is a viable solution to examine the potential effect of the exposure of manuscripts to reviewers at journals. Our exploratory study is a mixture of quantitative and qualitative methods and aims to contribute to this new stream of research on preprints by reconstructing 6,024 preprint-publication pairs across multiple fields and examining changes between versions.

### 1.1 Background

Recent studies on the effect of peer review on manuscripts suggest that peer review, at least in the social sciences, disproportionately focuses on the theoretical framing of studies, rather than their methodological content (Strang & Siler, 2015; Teplitskiy, 2016). Strang and Siler (2015), through qualitatively analysing 38 organisational studies articles, found that the theoretical framing of studies and the interpretation of results were most altered during peer review, while methodologies were largely unchanged. Reference lists grew on average by 26% between submitted and published versions, however references were both added and removed as authors revised their interpretations of results in line with reviewer criticism.

Teplitskiy (2016) compared 30 pairs of manuscripts in quantitative sociology before and after peer review and found that the manuscripts predominantly changed in their theoretical framing, rather than the methodology. These changes were applied to embed results in a theoretical framework and better convey a relation to the body of literature in sociology (Teplitskiy, 2016), although these studies were methodologically sound and publishable without such revisions (Teplitskiy et al., 2018). However, an analysis of *arXiv* and *bioRxiv* preprints and their published versions found little changes in the titles, abstracts or manuscript text between versions of a sample of physics and biology manuscripts (Klein et al., 2019). This would suggest that the focus on theoretical reframing is not shared by reviewers from physics and biology.

Studies in the medical sciences have found that peer review modestly improved the overall manuscript quality, in particular the generalisability and validity of results (Goodman et al., 1994), slightly improved readability, and increased the median length of articles by 2.6% (Roberts et al., 1994). However, peer review was found to be ineffective in detecting deficient reporting of methods and results in accordance with CONSORT clinical trial reporting guidelines (Hopewell et al., 2014). Carneiro et al. (2020) similarly found a marginal increase of reporting quality in their sample of *bioRxiv* preprints and associated publications, and 27% of them even decreased in reporting quality between versions.

In addition to changes suggested by well-intentioned reviewers, unethical review practices can also influence manuscripts. Such practices include coercive citations wherein editors or reviewers unnecessarily request citations to their own work or work previously published in these journals. Although the estimated rate of these practices varies from field to field (Carneiro et al., 2020; Wilhite & Fong, 2012), a study found that nearly a third of all citations recommended by reviewers were to their own works (Thombs et al., 2015). As such, reference changes during peer review may reflect a degree of unscrupulous peer review practices, in addition to legitimate revisions.

To investigate how peer review influences manuscripts and whether differences in its effect between disciplines could be identified, we matched 6,024 preprints across multiple disciplines to their published forms. We then quantified the changes made to reference lists during peer review, and identified the sections of publications that underwent the most extensive referencing changes. We validated our results through manual examination of the reference list changes and, via a qualitative analysis of the context of referencing changes, we reflected on the potential reasons for the changes. We concentrated on changes in reference lists between document versions as changes in references reveal reconstructions in the foundations on which authors have built their arguments and represent their position in the academic community. Considering multiple fields is crucial to acknowledge the different methods and foci of disciplines, especially as many studies of the effect of peer review have focused on the social and medical sciences (Carneiro et al., 2020; Goodman et al., 1994; Hopewell et al., 2014; Roberts et al., 1994; Strang & Siler, 2015; Teplitskiy, 2016).

There are limitations to using references as the measurement of the effect of peer review. For instance, changes may be made to the manuscript which do not affect referencing and these changes will not be captured through our method. However, Strang and Siler (2015) found moderate correlations between the intensity of peer review critiques and the extent of bibliographic changes made between manuscript versions, with more intense criticism typically associated with greater bibliographic change ($r$ = 0.31-0.59). These findings suggest that reference changes might be a suitable proxy for the effect of peer review on manuscripts.

## 2. Materials and Methods

We took three approaches to examining preprint-publication differences in this study: 1) a broad descriptive and quantitative view of all pairs and their changes in reference lists by discipline, 2) a more fine-grained view of publications to examine references' location in the full-text, in addition to reference list changes, and 3) a manual *gold standard* analysis of a subset of our data to assess the robustness of our conclusions.

For the first approach, as shown in Figure 1, we first identified all preprints indexed in *Dimensions* (Herzog et al., 2020) using the in-house version of data up to 26 April 2019 maintained by the German Competence Centre for Bibliometrics (KB). We selected *Dimensions* because it is the largest bibliometric database currently available (Visser et al. 2020) and, unlike other bibliometric databases, *Dimensions* includes preprints from popular servers such as *arXiv*, *medRxiv*, *bioRxiv*, *SSRN* and *OSF*. *Dimensions'* large size increased the likelihood of the published version being indexed in the database, expanding the possible sample size. Further, by using reference data for both preprints and publications from *Dimensions*, we could use internal identifiers to match references between lists rather than recreating this matching using metadata, which can be more error-prone. However, there are some limitations to *Dimensions'* use: *Dimensions* covers only a subset of all the preprints in repositories and its coverage of their reference lists is not complete. Furthermore, *Dimensions*, like other bibliometric databases, has a biased coverage of different disciplines, e.g., lower coverage for social sciences and humanities. Also, similar to other bibliometric databases, items not yet disambiguated in *Dimensions* are not included in the reference lists, which affects the completeness of coverage of both preprint and published reference lists. Furthermore, depositing a preprint on a server prior to publication is highly skewed toward specific disciplines and there are still disciplinary journals that do not allow authors to post preprint versions of manuscripts submitted to their journal. Nevertheless, Dimensions is the only large-scale bibliometric database indexing preprints so far (Herzog et al., 2020), as Scopus has only recently started indexing preprints.

Our search of *Dimensions* returned 373,563 preprints deposited to repositories between 2000 and 2018, of which 25,032 had reference lists. We then matched these preprints to publications indexed in the KB's version of Clarivate's *Web of Science* (WOS) that were published in the same year or subsequent two years based on a Jaro Winkler similarity between titles of more than 80%. We conducted the matching using the *stringdist package* in R (Loo et al., 2020) and identified a total of 2,986 pairs. We used a window of two years after the preprint's release to reduce the likelihood of false matches, given that 90% of *bioRxiv* preprints (as an example) are published within one year (Abdill & Blekhman, 2019). We used titles as the matching condition as prior studies have found titles rarely varied between preprint and publication versions (Klein et al., 2019).

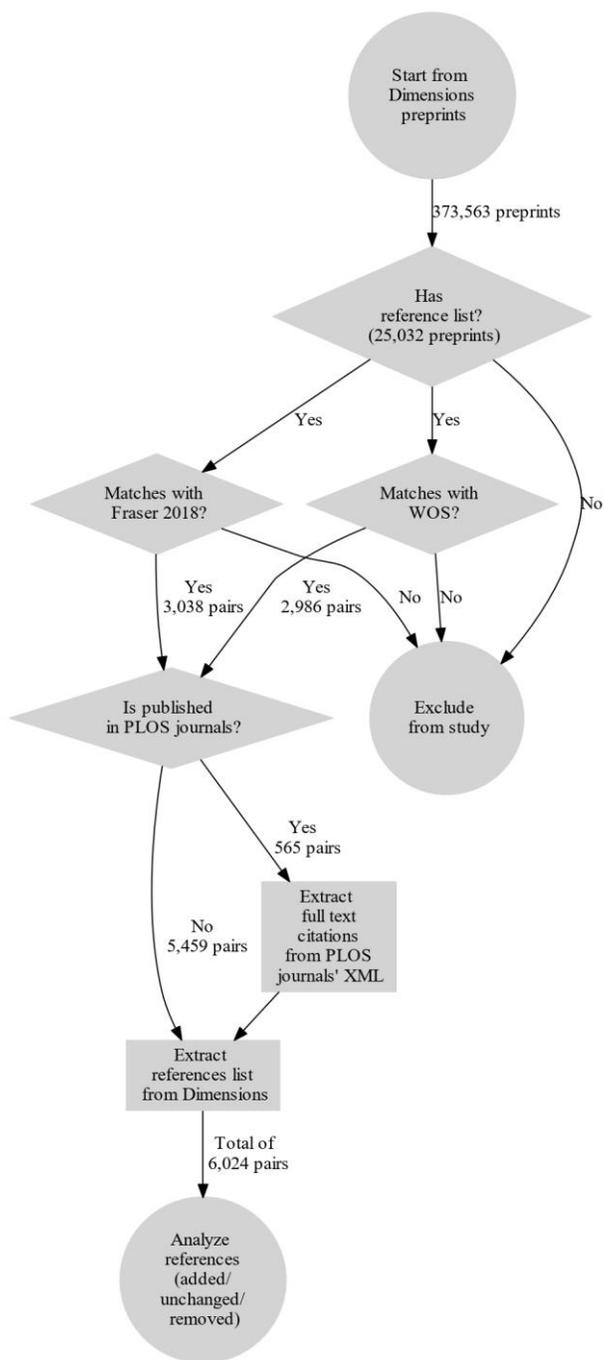

*Figure 1: Preprint and publication matching process (final chosen number of pairs: 6,024)*

To complement our data, we included a set of 3,038 preprint-publication pairs identified and published by Fraser et al. (2019). They matched preprints submitted to *bioRxiv* between November 2013 and December 2017 to publications by first querying the DOIs in *Crossref*'s "relationship" property via the API and then scraping the *bioRxiv* website for publication notices. They also applied fuzzy matching to *Scopus* publications based on author names, title, and the first 100 characters of the abstract, using the Jaro Winkler distance with a threshold of more than 80% similarity (Fraser et al., 2019). This additional dataset extended our coverage from *WOS* to also *Scopus*.

For all matched pairs from both sources (WOS and Fraser sets), we obtained the publications' DOIs, extracted the corresponding reference lists from *Dimensions,* and matched the references between lists based on *Dimensions'* internal identifiers. Once we had compiled all the reference data, we merged and de-duplicated the pair sets from both sources, resulting in a final sample of 6,024 pairs.

We then confronted the preprint reference list with the published version. Following Strang and Siler's (2015) methodology, we assigned each reference as *unchanged*, *added*, or *removed*. *Unchanged* references were present in both reference lists. References cited in the preprint but not cited in the publication were *removed*, while references cited in the publication but not the preprint were *added*. For each pair, we then calculated the proportion of references of each status based on the total number of unique references in the pair's combined reference lists (for an example, see Fig. S1 in the Supplementary Material).

Our second approach analysed which sections of publications were most affected by revisions in each pair. We used 565 document pairs from our sample that had published versions in *PLOS* journals because this publisher provides full-texts in XML documents that include hyperlinks to cited references, allowing us to track in which publication sections each reference was cited. We downloaded the full *PLOS* corpus in November 2019 and extracted the full-texts of publications in our matched pairs based on the publication's DOI. We first standardised the section names to *Introduction*, *Method*, *Results*, *Discussion and Conclusions*, and *Supplementary Material* and then calculated the change in sections based on the proportion of references added or unchanged. Our section analysis was limited to added and unchanged references as we did not have the full-text of preprints because, although some preprint servers such as ArXiv provide the full-text of only LaTeX-based preprints, they are often highly unstructured depending on the writing tools authors used. Without the preprint full-text, we could not identify the sections of preprints from which references were removed. In this analysis, the reference status (e.g., added or unchanged) comes from comparing the preprint and publications' reference lists and the section(s) comes from analysing the publications' full-text XML files. We assigned cases to an "Unknown" section when in-text citations were not clearly assigned to a specific section. Finally, we standardised the journal names in each reference and matched them to the journal in which the published version appeared to investigate citations to the publishing journal.

In investigating disciplinary differences, we used the OECD's Fields of Science and Technology (FOS) classification (OECD, 2007). The FOS is a two-level classification of 42 disciplines that aggregate into six major fields: *Agricultural Sciences* (AS), *Engineering and Technology* (ET), *Natural Sciences* (NS), *Medical and Health Sciences* (MHS), *Humanities* (H) and *Social Sciences* (SS). We mapped the pairs to the FOS classification based on the publication's classifications in *WOS* or *Scopus* and WOS/Scopus-to-FOS correspondences provided by *Clarivate Analytics* and *Elsevier*. Note that some publications were assigned to multiple fields and so our results used multiple counting of publications to each assigned discipline.

Our third approach aimed to validate our matching process and the *Dimensions*' reference list data by undertaking a manual validation process for a sample of our 6,024 pairs. We compared the data in the preprint repositories and journal websites against our data for a random sample of 10 pairs from each of the 5 fields and 2 sources (*WOS* and Fraser sets), and all 5 pairs from the humanities contained in our dataset. We also included an additional 20 pairs from publications in *PLOS* journals to validate our analysis of changes by section. Thus, our final sample for manual checks consisted of 125 pairs.

We conducted several checks for accuracy using this sample. First, we compared the preprint and publication titles to assess the accuracy of matching the preprint-publication pairs. For correct matches, we then accessed the reference lists of the preprints and publications in their online versions. We manually compared the online versions of the pairs' reference lists against the lists retrieved through *Dimensions* to identify the presence or absence of references in the *Dimensions* data. We also extracted the document type and publication year for each reference from the online versions. From these checks, we determined the accuracy of matching the pairs, their reference lists and the completeness of the *Dimensions* reference lists. Then, based on the complete reference lists obtained online, we calculated the actual status of references as added, removed or unchanged between versions, and compared this with our results from the machine-based process to assess its validity. We also identified the section of the document – standardised as previously described – in which the added or removed references were cited.

Finally, to supplement our quantitative analysis of referencing changes, we qualitatively examined potential reasons for changes in references using the randomly sampled pairs. For each added reference, one of the authors (DS) identified the reference's in-text citation in the publication and the corresponding paragraph in the preprint, but which did not contain the reference (and vice versa for the removed references). DS then examined the surrounding text for changes and assigned a reason for adding the reference based on the text changes and the type of reference added. DS took an inductive coding approach (Elo & Kyngäs, 2008) to coding the reasons for referencing changes. Once DS had coded all reference changes, she organised the codes into broader categories reflecting the primary themes under which the changes occurred.

In most cases, there was relatively little difference in the structure and text of the documents and the corresponding paragraphs in each version were easily identifiable. In a small number of cases, authors had rewritten entire introductions or discussions. In these cases, the reason for change was assigned to the category "Change in background framing" or "Change in results interpretation" based on whether the introduction or discussion was rewritten. Most changed references were cited in the text only once, but where they occurred multiple times, one reason was assigned for each change. Coding and this data analysis was carried out in Microsoft Excel 2013. None of the authors in the sampled manuscripts were personally known to DS.

## 3. Results

We first present the results of our third analysis, the validation study, as these results are important for interpreting the results of the first (quantitative) and second (full-text) analyses. Of the 125 pairs selected for manual validation, comparing the preprint and publication titles confirmed that 113 (90.4%) were correct matches. After removing the 12 (9.6%) false matches and 15 pairs inaccessible due to the removal of the preprint from its repository, we analysed 98 pairs. By comparing the 98 pairs' online reference lists against the *Dimensions* lists, we identified that, on average, 30.8% of references were missing from the *Dimensions* reference lists. However, the reference lists of preprints were much more incomplete (56.4% references missing) than were the publication's reference lists (10.2%). References missing from the publication lists tended to be older publications, books, software, reports and non-English language documents, while preprints' missing references included such documents but also a large number of recent, English-language publications that were present in the publication reference list data.

Overall, for the 6,228 references in the 98 pairs, there were 495 (7.9%) *added* references, 225 (3.6%) *removed* references, and 5,443 (87.4%) were *unchanged* between the preprint and publication (65, 1.0%, were duplicate references or incorrectly included in a reference list). Table 1 shows the number and percentage of references in the 98 pairs by their status derived in the quantitative analysis and their actual status assigned in the validation process. Notably here, the majority (2,569, 84.6%) of references considered *added* were actually *unchanged*, driven by the missing preprint references. Of the references initially considered *removed*, validation confirmed 56.1% (87) were accurate, while 30.3% (47) were actually *unchanged.* Nearly all (2,329, 99.2%) of the *unchanged* references were accurate, however, the total number of *unchanged* references is under-reported in the initial status due to the inaccurate inclusion of many *unchanged* references as *added* or *removed*.

Table 1: Number and percentage of references in validation sample by derived status and actual status. Incorrect listing describes references incorrectly included in one reference list.

| | Actual status | | | | |
| --- | --- | --- | --- | --- | --- |
| **Derived status** | **Added (%)** | **Removed (%)** | **Unchanged (%)** | **Incorrect listing (%)** | **Total (%)** |
| Added | 430 (14.2) | 0 (0.0) | 2,569 (84.6) | 36 (1.2) | 3,305 (100.0) |
| Removed | 0 (0.0) | 87 (56.1) | 47 (30.3) | 21 (13.5) | 155 (100.0) |
| Unchanged | 4 (0.2) | 7 (0.3) | 2,329 (99.2) | 8 (0.3) | 2,348 (100.0) |
| Missing | 61 (8.8) | 131 (19.0) | 498 (72.2) | 0 (0.0) | 690 (100.0) |
| Total | 495 (7.9) | 225 (3.6) | 5,443 (87.4) | 65 (1.0) | 6,228 (100.0) |

Figure 2 compares the proportion of references *added, removed* and *unchanged* by the FOS field. The top panel shows the macro view based on the total sample and the bottom panel shows the results of our manual *gold standard* validation. Humanities is not present in the validation set due to inaccessibility. We found that the validated proportion of *unchanged* references in all fields is higher and the proportions of *added* references much lower than in our total sample. In general however, the distributions of *unchanged* and *removed* references in our total and validation samples are in agreement. Our analysis of reference changes by manuscript section using the gold standard subset also validated the patterns observed for the total sample (see Fig. S3 in the Supplementary Material).

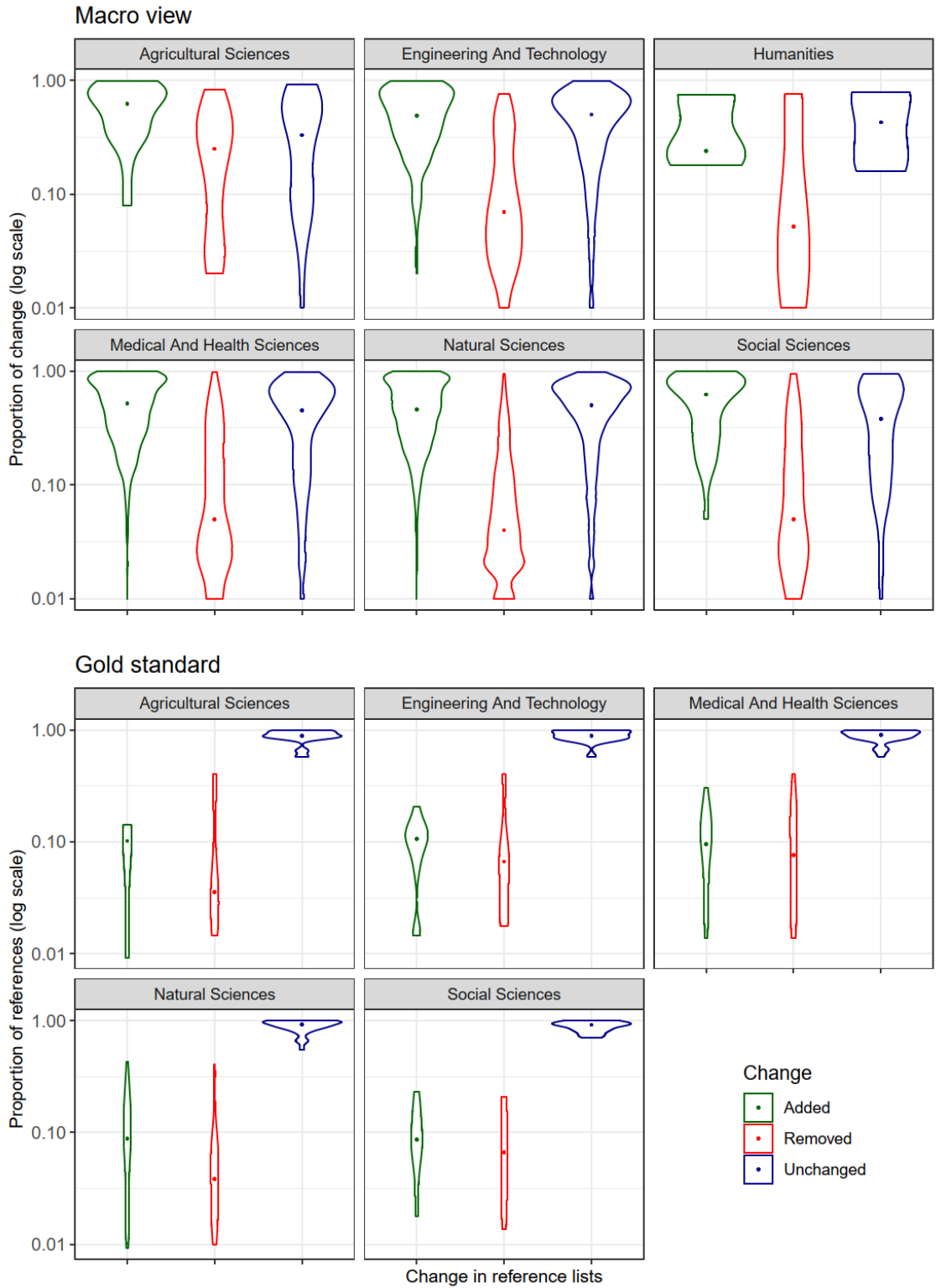

*Figure 2: Distribution of references added, removed or unchanged in macro quantitative view (top) and validation sample (bottom) by field (dots indicate median)*

Based on our macro quantitative analysis of all 6,024 pairs, Figure 3 shows the general trend of referencing changes and highlights disciplinary differences. We found outlying pairs with higher proportions of removed references in many disciplines (e.g., 75% in some cases), but the general trend was that most disciplines had a median of <25% *removed* references. However, there were ten disciplines with median *removed* references close to or >25% (see Table S1 in the Supplementary Material for details). Importantly, as *Dimensions*' coverage of preprint references was less reliable compared to publications' references, the proportion of *added* references are inflated here and part of these references would be *unchanged* if *Dimensions*' coverage was complete, so these data should be interpreted cautiously. Also, some fields did not have any *unchanged* references which can further point to problems with *Dimensions'* coverage of preprint reference lists.

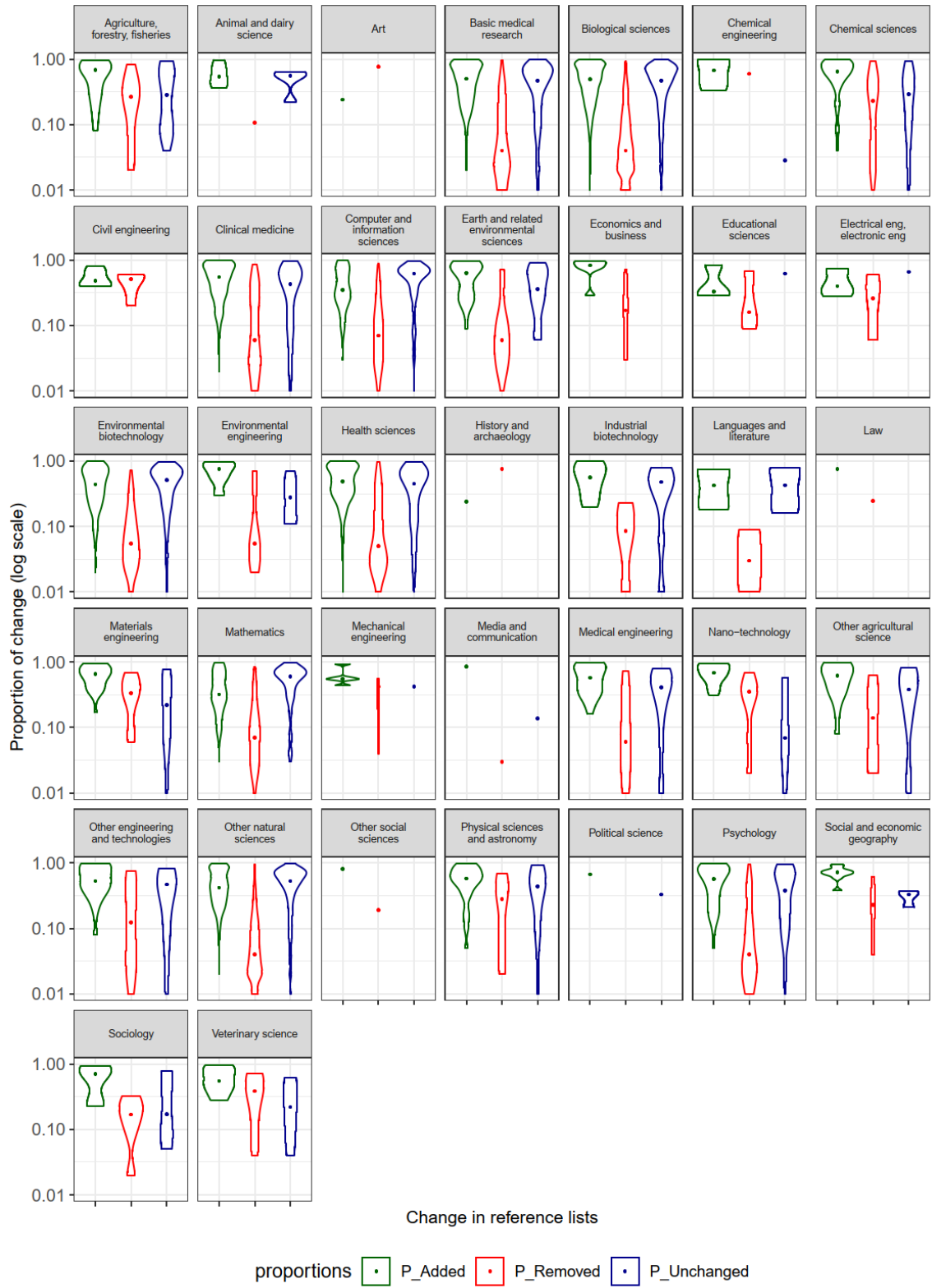

*Figure 3: Distribution of proportion of references added, removed or unchanged by WoS discipline. Dots in the centres of violins show the median (dots indicate median).*

Figure 4 presents the results of our second analysis examining which sections of manuscripts undergo the most referencing changes during peer review. The top panel shows the proportions of references added or unchanged by manuscript section using full-texts from *PLOS* journals. Here, the proportions are relative to all references in the single publication then aggregated over journals to allow comparison between sections and journals. In addition, "all_journals" show the aggregate trend of all of these journals. *PLOS* journals on the Y axis are sorted based on the decreasing number of pairs in our sample. *PLOS Genetics*, *Biology*, *Computational Biology*, and *One* all have a primary focus in the natural sciences, while *PLOS Pathogens* is a mix of natural and medical sciences, and *Neglected Tropical Diseases* (NTD) is primarily a medical sciences journal. Of course, PLOS One has a relatively small share (21%[1]) of social science publications.

Overall, in nearly all sections and journals, the share of *added* references was higher than *unchanged*, and this difference is highest in the medical science journals. However, this is in part due to missing references in *Dimensions'* preprints. Our comparison of sections indicates that in all journals, the majority of bibliographic changes occurred in the introductions and discussions of the manuscripts. However, as the introduction and discussions are the sections used to frame the study in existing literature, it is more likely that references would be added here than in the methodology or results (Bertin et al., 2016). As such, the bottom panel of Figure 3 shows the ratio of *added* to *unchanged* references in each section to account for the uneven distribution of references across a manuscript. Here, in 3 of the 4 natural science journals (i.e., *PLOS One*, *Computational Biology* and *Biology*), the methods section underwent the most bibliographic change, with around two times as many references *added* as *unchanged*, while in the more medically-oriented journals (i.e., NTD and *Pathogens*), there were greater change in the results and discussion sections. See Table S2 in the Supplementary Material for the full data for this Figure.

We found no notable differences in the proportions of references to the publishing journal that were added, removed, or unchanged after peer review between fields or signs of coercive citations (see Fig. S2 in the Supplementary Material).

---

[1] Based on count of all PLOS One publications (260,009) versus those indicated as social sciences (56,196) as of December 7th 2021 accessed in https://journals.plos.org/plosone/

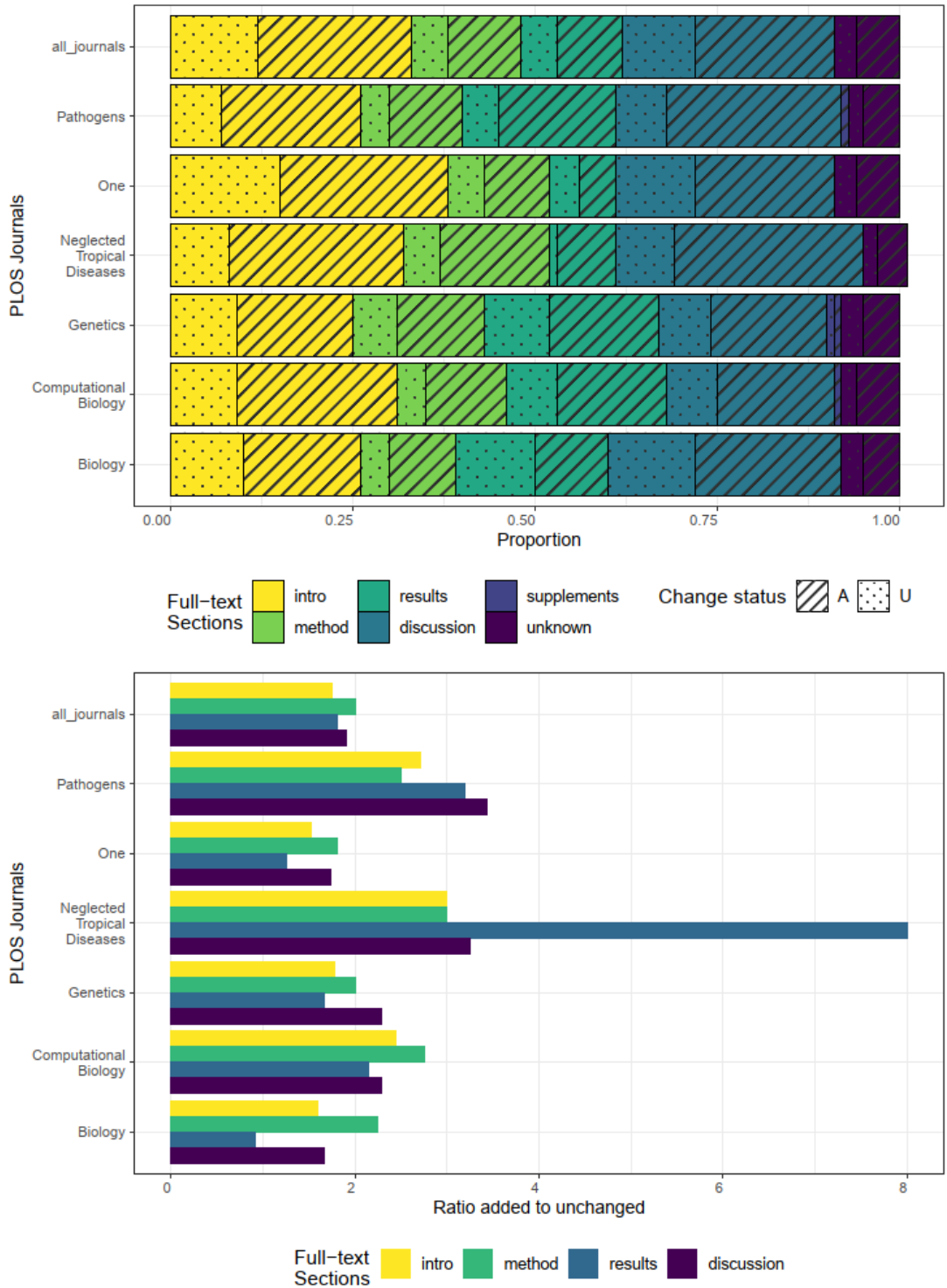

*Figure 4: Distribution of references change in full text sections by PLOS journals (top) and added to unchanged proportion by section (bottom, U = unchanged, A = added, all_journals = aggregate trend of all PLOS journals)*

Finally, we present the qualitative results from our analysis of the 98 pairs to identify themes under which references were changed during peer review. The number and percentage of references changed in accordance with each of the ten identified themes by field is shown in Table 2. Note that some references are included in more than one field and so the sum of the fields is greater than the total.

Four themes were associated with notable changes to the manuscript's content between versions. The most common of these themes was changing the interpretation of results, which resulted in removing 8.2% (19) of references and adding 25.3% (133). The majority of references were added when authors included additional text to better interpret their results in regard to existing literature, or addressed limitations or the significance of the study. This occurred particularly often in the social sciences, where it accounted for 32.5% (27) of all *added* references. However, the small percentage of *removed* references (3, 4.4%) compared to *added* in the social sciences, and also the agricultural sciences and engineering fields, suggests the results were not completely reframed but were instead more thoroughly contextualised within the existing literature. In comparison, in the natural sciences, changes in interpreting results prompted nearly equal percentages of *added* and *removed* references, and there was an elevated percentage of *removed* references in the medical sciences also, suggesting results are more extensively reframed within the literature in these fields.

References were also commonly added to expand the background information about the study subject (130, 24.7%). In the majority of such cases (82, 63.1%), references were added to establish a theoretical or practical foundation for the study based on the existing literature. These changes typically occurred in the introduction sections, and authors also removed 16.5% (38) of references in this revision process. This was the key theme for adding references for the natural, medical, and agricultural sciences. Again, the natural and medical sciences had higher levels of references removed for this purpose, alongside the added references, suggesting manuscripts in these fields undergo more transformation in the theoretical foundations of these studies than in other fields.

Changes in the study's methodology accounted for 15.4% (81) of *added* references and 5.2% (12) of *removed* references. Here, authors added references to provide missing or more extensive detail about existing data or processes (29 references, 35.8%), to justify their selection of particular methods (17, 21.0%), or to provide references for specific methods (9, 11.1%). Another set of reference changes were triggered by changes in the study's methodology. Twenty-six references (32.1%) were added because new analyses were conducted, and 12 references (5.2%) were removed as authors changed the study's method and removed associated references. Such methodological changes may have been requested by reviewers or added of the authors' own accord before submitting the manuscript to the journal. Engineering papers appeared to undergo the most methodological changes, with higher percentages of references added and, in particular, removed for this purpose than in other fields.

Authors added another small set of references (20, 3.8%) when discussing new suggestions for applications of the study's findings and directions of future inquiry. Like additions made to address limitations or methodological issues, these sections regarding future directions often appeared to have been specifically requested by reviewers as they frequently appeared as wholly new paragraphs sandwiched between sections with no change. However, this is only our impression, as we did not have access to the reviewers' reports to confirm the requests. The percentage of references added for this purpose was relatively equal across fields (2.8-4.8%).

The remaining six themes represented referencing changes that reflected small updates to the manuscript. These include adding references for previously uncited software or other tools used (24, 4.6%). References were also added to support previously unsupported knowledge claims (27, 5.1%). The social and medical sciences had substantially fewer additions for this reason than the other fields (1-4, 1.2-1.8% compared to 4-16, 5.2-10.3%). References were also added (45, 8.6%) or removed (64, 27.7%) without making changes to the surrounding text. This theme was the most common circumstance under which references were removed in all fields. Also, a small percentage of references were removed (16, 6.9%) and added (7, 1.3%) as authors updated references from citing, for instance, a conference paper in the preprint to citing its published version in the publication. This was notably more common in engineering and the natural sciences than other fields. Finally, referencing mistakes, in which a reference was included in the reference list but not cited in the document's text, explained over a third of removals (82, 35.5%). We show in Fig. S4 of the Supplementary Material the percentage of references removed and added in each theme by manuscript section.

Table 2: Number and percentage of references by thematic group and field

| Status | Reason | Soc. sci. (%) | Nat. sci. (%) | Med. sci. (%) | Agr. sci. (%) | Engin. (%) | Total (%) |
|---|---|---|---|---|---|---|---|
| Added | Change in results interpretation | 27 (32.5) | 66 (21.3) | 62 (28.1) | 12 (20.7) | 11 (15.5) | 133 (25.3) |
| | Change in background framing | 8 (9.6) | 87 (28.1) | 67 (30.3) | 18 (31.0) | 6 (8.5) | 130 (24.7) |
| | Change in methodological details | 10 (12.0) | 52 (16.8) | 27 (12.2) | 8 (13.8) | 13 (18.3) | 81 (15.4) |
| | Referencing mistakes | 28 (33.7) | 24 (7.7) | 32 (14.5) | 1 (1.7) | 16 (22.5) | 57 (10.8) |
| | Change without modifying text | 4 (4.8) | 34 (11.0) | 16 (7.2) | 3 (5.2) | 4 (5.6) | 45 (8.6) |
| | Support for unsupported claims | 1 (1.2) | 16 (5.2) | 4 (1.8) | 6 (10.3) | 4 (5.6) | 27 (5.1) |
| | References for software, tools, etc | 1 (1.2) | 15 (4.8) | 5 (2.3) | 6 (10.3) | 13 (18.3) | 24 (4.6) |
| | Future directions | 4 (4.8) | 9 (2.9) | 7 (3.2) | 2 (3.4) | 2 (2.8) | 20 (3.8) |
| | Updated references | 0 (0.0) | 6 (1.9) | 1 (0.5) | 1 (1.7) | 2 (2.8) | 7 (1.3) |
| | Other | 0 (0.0) | 1 (0.3) | 0 (0.0) | 1 (1.7) | 0 (0.0) | 2 (0.4) |
| | **Total added** | **83 (100.0)** | **310 (100.0)** | **221 (100.0)** | **58 (100.0)** | **71 (100.0)** | **526 (100.0)** |
| Removed | Change in results interpretation | 3 (4.4) | 28 (20.6) | 14 (10.9) | 3 (5.2) | 0 (0.0) | 19 (8.2) |
| | Change in background framing | 3 (4.4) | 11 (8.1) | 12 (9.4) | 0 (0.0) | 1 (2.1) | 38 (16.5) |
| | Change in methodological details | 0 (0.0) | 8 (5,9) | 3 (2.4) | 4 (6,9) | 8 (16,7) | 12 (5.2) |
| | Updated references | 2 (2,9) | 16 (11,8) | 3 (2,4) | 4 (6,9) | 8 (16,7) | 16 (6.9) |
| | Change without modifying text | 6 (8,8) | 54 (39,7) | 31 (24,6) | 42 (72,4) | 27 (56,3) | 64 (27.7) |
| | Referencing mistakes | 54 (79,4) | 19 (14.0) | 63 (50.0) | 5 (8,6) | 4 (8,3) | 82 (35.5) |
| | **Total removed** | **68 (100.0)** | **136 (100.0)** | **128 (100.0)** | **58 (100.0)** | **48 (100.0)** | **231 (100.0)** |
| **Pairs** | | **19** | **62** | **43** | **14** | **18** | **98** |
| **References** | | **701** | **2,573** | **1,841** | **492** | **509** | **6,163** |

## 4. Discussion

In this study, we explored how manuscripts are revised during their development, i.e., from preprint to publication, as measured by referencing changes. Although we estimated revisions and changes of references only indirectly by comparing input (i.e., preprints) and output (publications), our study helps to improve our understanding of the effect of peer review on manuscript development (Batagelj et al., 2017; Garcıa-Costa et al., 2021). We could not reconstruct the motivations behind authors' citing behaviour and point-to-point adaptations to reviewer requests, but nevertheless, our results offer insight into how peer review and revisions, or anticipatory modifications of authors, change the reference lists of manuscripts.

In our quantitative investigation of 6,024 preprint-publication pairs, we considered manuscripts from ten disciplines, mainly from the natural sciences, with a median of up to 25% or in some cases even 50% of references removed between the preprint and publication. Our in-depth examination of the full-texts of publications in *PLOS* journals showed that, in the natural sciences, methods sections are most bibliographically transformed, while referencing in the results and discussion sections changed most in the medical sciences. Furthermore, we found that publications in *pure* disciplines tended to change less, while inter/multidisciplinary publications (e.g., in *PLOS Genetics* and *Computational Biology*) undergo a mixture of changes similar to that of the fields they bridge.

However, overall, referencing between preprints and publications was notably stable. Nearly 90% of references present in the preprint were retained in the publication, while 8% were added, and only 4% were removed. Indeed, references were most commonly removed because they were incorrectly included in the preprint reference list and this was corrected in the publication, suggesting the publication process improves the accuracy of citation practices. We identified nine additional themes under which bibliographic changes occurred. Four of these pertained to changes that did not result in substantial or any changes to the manuscript's content, such as adding references for software or for claims that were unsupported in the preprint, and changing references without changing the surrounding text. This latter practice was particularly common for removing references and may perhaps relate to reductions related to word count limitations (Teplitskiy, 2016). Authors adding references for these purposes suggests reviewers have an important role in detecting unsupported claims and ensuring software, and methods are appropriately cited.

The bulk of referencing changes resulted from changes to the structure or content of the manuscript, including reframing the study's placement or interpretation of its results within existing literature, or changes in the methods. These changes align with the purposes of peer review to assess a study's methodological soundness and improve the communication of its results to the academic community (De Vries et al., 2009; Rigby et al., 2018). However, we found differences between fields in the probable effect of peer review. Referencing changes in engineering was more oriented toward methodological soundness, while in the other fields, the focus appeared to be on the theoretical framing of the study and its results. While the majority of reference additions in all of these fields occurred through reframing the study's theoretical background and results, we also observed higher levels of removing references for this purpose in the natural and medical sciences than in the social and agricultural sciences. This suggests that natural and medical science publications might undergo more extensive reframing of studies, with substitution of references as foundational and explanatory theories are exchanged, whereas studies in the social and agricultural sciences become more embedded in the field, with references added but not removed.

This reframing may reflect the negotiation between authors and reviewers in how studies and their results are interpreted (Teplitskiy, 2016). However, in the social sciences there is perhaps larger overlap in the theories applicable to results than in the natural and medical sciences (Watts, 2017), where interpretation under one theory may necessarily preclude interpretation under another, thus requiring the removal of existing and addition of new references (Flaherty, 2016).

Arguably, the reviewers' predominant focus on theory in most fields may reflect reviewers' awareness, as academics themselves, of the resource-intensiveness of the scientific process. For many reasons, a researcher usually cannot simply collect new data or run entirely new analyses to address reviewers' concerns, particularly should these concerns not pertain to fatal flaws. Further, given the pro-bono and ad hoc nature of peer review, reviewers may lack the resources in terms of time and availability of data to thoroughly analyse or reproduce a study's results. As such, reviewers may focus on addressing the disconnection between the theoretical framing and methodological aspects of the study by suggesting reframing of the theory to align with the questions that can be answered by the available data in a data-driven approach, rather than retaining the question and adjusting the data in a question-driven approach (Teplitskiy, 2016). However, confirming this hypothesis would require access to peer review data to analyse authors' and reviewers' correspondence, which is typically inaccessible (Squazzoni et al., 2020).

Finally, our study has certain limitations. While using references as a proxy for changes in a manuscript allowed us to examine a large number of manuscripts, the reliance solely on references might be flawed. For instance, methodologies, particularly when innovative, may not include references, providing unreliable results about the extent to which they are critiqued and addressed during peer review. Further, influences of peer review on manuscripts that do not affect referencing, such as improving the quality of communication, cannot be captured through our analyses.

Although we used two databases to increase the likelihood of retrieving complete reference lists, we note that incompleteness in reference lists is still a problem, particularly in the humanities and social sciences, and we are likely missing references. Through matching references from *Dimensions* to the reference lists in *WOS*, it is probable that we lost some references, which might decrease the magnitude of observed trends. Our results should then be considered on the lower end of the continuum. Matching preprints with their published versions was also difficult. We used multiple approaches, e.g., similarity between titles of both sides while complementing it with currently existing pairs of DOIs (Fraser et al., 2019), but this could still be improved.

Furthermore, our study only estimated the impact of peer review on manuscript change by considering the articles eventually published, ignoring rejected articles. This could be indirectly controlled by comparing our sample of preprints with similar ones that were not published, as authors may improve their manuscript beyond the advice given by reviewers. Thus, the observed trends in referencing changes cannot be solely attributed to peer review, but can include changes and improvements made by authors of their own accord. As such, we may have overestimated the effect of peer review on bibliographic changes.

Noting these limitations, our study offers useful insight into the referencing changes manuscripts undergo during peer review in different fields. In tracking manuscript changes, we see a two-fold function of 'developmental peer review': examining the methodological soundness of manuscripts while improving the communication of findings and their appropriate credit (Horbach & Halffman, 2018). Given that competition for priority has increased in academia and research has become more complex in terms of methodologies, instruments and tools, ensuring appropriate credit and finding cumulativeness by accurate referencing is of paramount importance, even only to avoid credit misattribution (Edwards & Roy, 2016). Although journals are now under pressure for rapid dissemination of innovative scientific findings, their real function is to ensure the rigour and validity of scientific claims through intensive collaboration of authors and reviewers in improving standards of quality (Kharasch et al., 2021).

Future studies could bridge the advantage of sample size and specificity in detection of changes by using automated textual analysis, as is used for plagiarism detection software, which could facilitate identifying changes in the text of a large sample of manuscripts (Strang & Dokshin, 2019). This would preclude the reliance on reference lists and the issues we encountered here but would involve dealing with complexities of unstructured textual data. Current initiatives pursuing the goal of opening up peer review data (Squazzoni et al., 2017, 2020) could help to examine the effect of peer review on manuscript changes more precisely, perhaps also looking at reference suggestions by reviewers (e.g., (Casnici, Grimaldo, Gilbert, & Squazzoni, 2017)). Whether those references that were removed were due to reducing the word count of the manuscript, could be an avenue for future studies by investigating journal guidelines for presence of strict word counts.

**Acknowledgements**

We would like to thank Nicholas Fraser, Fakhri Momeni, Philipp Mayr, and Isabella Peters for their publicly available data (Fraser et al., 2019) that we used to complement our dataset. We would also like to thank Martin Reinhart and Bahar Mehmani for helpful comments on an earlier version of this paper.

**Data Availability**

Micro publication level data cannot be made publicly available due to the licensing and contract terms of the original data provider. However, contact authors for preprint-publication pair level data

**Author contributions**

AA and DS have contributed equally. They designed the study, gathered, analysed and interpreted the data, and wrote the manuscript. FS interpreted the data and wrote the manuscript.

**Funding information**

Data was obtained from Kompetenzzentrum Bibliometrie (Competence Centre for Bibliometrics), Germany, which is funded by the Federal Ministry for Education and Research (BMBF), Germany, with grant number 01PQ17001. FS was supported by a grant of the MIUR-Italian Minister for Education, University and Research (20178TRM3F_002) and a grant from the University of Milan (PSR2015-17 Transition Grant).

**Declaration of competing interest**

The author(s) declare no competing interests.

Supplementary materials for

# A study of referencing changes in preprint-publication pairs across multiple fields

**Aliakbar Akbaritabar**[*1¶], **Dimity Stephen**[2¶], **Flaminio Squazzoni**[3]

[*1] Laboratory of Digital and Computational Demography, Max Planck Institute for Demographic Research (MPIDR), Rostock, Germany; ORCID = 0000-0003-3828-1533

[2] German Centre for Higher Education Research and Science Studies (DZHW), Berlin, Germany; ORCID = 0000-0002-7787-6081

[3] Department of Social and Political Sciences, University of Milan, Milan, Italy; ORCID = 0000-0002-6503-6077

* Corresponding author

E-mail: akbaritabar@demogr.mpg.de; akbaritabar@gmail.com (AA)

¶These authors contributed equally to this work

Figure S1 shows an example of how we calculated the proportions of references added, removed or unchanged. We assumed that the set of references used in the preprint and/or publication were the main corpora of literature used by authors. Thus, in calculating the proportion of changes in reference lists, we considered these corpora as our baseline and we did not use only the reference list of published version. For instance, if a preprint had five references and two were removed, three were unchanged and one new reference was added, we calculated the proportions on the total six references in both preprint and publication.

| Preprint | Publication | Status |
|---|---|---|
| Ref. 1 | Ref. 1 | Unchanged |
| Ref. 2 | Ref. 2 | Unchanged |
| Ref. 3 | -- | Removed |
| Ref. 4 | Ref. 4 | Unchanged |
| Ref. 5 | -- | Removed |
| -- | Ref. 6 | Added |

| Status | Count | Proportion |
|---|---|---|
| Unchanged | 3 | 0.50 |
| Removed | 2 | 0.33 |
| Added | 1 | 0.17 |

*Figure S1: Method for determining preprint and publication references' change status*

Table S1 presents the aggregated version of data behind the Figure 1 regarding number of preprint-publication pairs, average number and average proportion of added, removed and unchanged references in FOS disciplines.

Table S1: Preprint and publication pairs by WOS disciplines while multiple counting (N = number of references, P = proportion of references)

| FOS | n_pairs | N_Unchanged | N_Removed | N_Added | P_Unchanged | P_Removed | P_Added |
|---|---|---|---|---|---|---|---|
| Biological sciences | 2,514 | 27.18 | 6.07 | 31.22 | 0.47 | 0.09 | 0.51 |
| Other natural sciences | 1,575 | 27.31 | 5.04 | 26.99 | 0.50 | 0.08 | 0.48 |
| Basic medical research | 701 | 28.05 | 7.93 | 34.50 | 0.45 | 0.11 | 0.52 |
| Clinical medicine | 430 | 24.51 | 10.19 | 32.46 | 0.42 | 0.17 | 0.55 |
| Health sciences | 360 | 19.54 | 7.96 | 25.16 | 0.44 | 0.14 | 0.53 |
| Environmental biotechnology | 341 | 21.46 | 5.05 | 22.76 | 0.49 | 0.10 | 0.48 |
| Psychology | 166 | 27.73 | 7.81 | 37.39 | 0.43 | 0.12 | 0.55 |
| Computer and information sciences | 152 | 19.16 | 5.19 | 15.95 | 0.54 | 0.13 | 0.45 |
| Mathematics | 128 | 17.03 | 5.00 | 13.83 | 0.54 | 0.13 | 0.43 |
| Chemical sciences | 83 | 16.35 | 14.10 | 35.67 | 0.33 | 0.27 | 0.63 |
| Earth and related environmental sciences | 56 | 25.58 | 9.07 | 30.18 | 0.44 | 0.13 | 0.58 |
| Physical sciences and astronomy | 40 | 23.84 | 16.44 | 32.49 | 0.44 | 0.28 | 0.57 |
| Medical engineering | 36 | 21.23 | 10.50 | 35.14 | 0.39 | 0.17 | 0.59 |
| Agriculture, forestry, fisheries | 31 | 18.75 | 17.12 | 40.23 | 0.37 | 0.28 | 0.58 |
| Materials engineering | 28 | 14.36 | 20.00 | 28.79 | 0.31 | 0.35 | 0.63 |
| Other engineering and technologies | 26 | 14.47 | 12.12 | 30.81 | 0.39 | 0.24 | 0.60 |
| Other agricultural science | 15 | 13.55 | 12.22 | 28.80 | 0.39 | 0.20 | 0.59 |
| Nano-technology | 14 | 6.20 | 21.86 | 33.21 | 0.21 | 0.31 | 0.69 |
| Environmental engineering | 12 | 17.29 | 10.88 | 33.17 | 0.32 | 0.18 | 0.70 |
| Veterinary science | 11 | 15.60 | 26.17 | 35.91 | 0.29 | 0.37 | 0.61 |
| Social and economic geography | 10 | 12.00 | 20.44 | 32.20 | 0.30 | 0.22 | 0.71 |
| Industrial biotechnology | 9 | 27.38 | 8.25 | 39.11 | 0.41 | 0.10 | 0.59 |
| Sociology | 9 | 19.43 | 9.17 | 36.78 | 0.34 | 0.17 | 0.62 |
| Economics and business | 7 | | 8.00 | 33.71 | | 0.21 | 0.79 |
| Animal and dairy science | 6 | 12.75 | 13.00 | 23.33 | 0.50 | 0.20 | 0.60 |
| Civil engineering | 6 | | 24.00 | 24.67 | | 0.46 | 0.54 |
| Mechanical engineering | 5 | 10.00 | 12.40 | 23.00 | 0.42 | 0.31 | 0.61 |
| Chemical engineering | 4 | 2.00 | 28.00 | 41.25 | 0.03 | 0.60 | 0.68 |
| Languages and literature | 4 | 32.75 | 2.67 | 31.25 | 0.48 | 0.04 | 0.48 |
| Educational sciences | 3 | 35.00 | 14.33 | 19.33 | 0.62 | 0.31 | 0.49 |
| Electrical eng, electronic eng | 3 | 54.00 | 16.67 | 24.67 | 0.66 | 0.31 | 0.47 |
| Law | 2 | | 9.00 | 27.00 | | 0.24 | 0.76 |
| Media and communication | 2 | 4.00 | 1.00 | 26.50 | 0.14 | 0.03 | 0.85 |
| Art | 1 | | 16.00 | 5.00 | | 0.76 | 0.24 |
| History and archaeology | 1 | | 16.00 | 5.00 | | 0.76 | 0.24 |
| Other social sciences | 1 | | 18.00 | 79.00 | | 0.19 | 0.81 |
| Political science | 1 | 6.00 | | 12.00 | 0.33 | | 0.67 |

Table S2 presents the data used in Figure 2. It presents number of preprint-publication pairs in each PLOS journal and proportion of changes in references used in full-text sections.

Table S2: References change in full text sections by PLOS journals (U = unchanged, A = added)

| PLOS journals | n_pairs | introA | introU | methodA | methodU | resultsA | resultsU | discussionA | discussionU | supplmentsA | supplmentsU | unknownA | unknownU |
|---|---|---|---|---|---|---|---|---|---|---|---|---|---|
| One | 306 | 0.23 | 0.15 | 0.09 | 0.05 | 0.05 | 0.04 | 0.19 | 0.11 | 0.00 | 0.00 | 0.06 | 0.03 |
| Computational Biology | 94 | 0.22 | 0.09 | 0.11 | 0.04 | 0.15 | 0.07 | 0.16 | 0.07 | 0.01 | 0.00 | 0.06 | 0.02 |
| Genetics | 83 | 0.16 | 0.09 | 0.12 | 0.06 | 0.15 | 0.09 | 0.16 | 0.07 | 0.01 | 0.01 | 0.05 | 0.03 |
| Pathogens | 31 | 0.19 | 0.07 | 0.10 | 0.04 | 0.16 | 0.05 | 0.24 | 0.07 | 0.01 | 0.00 | 0.05 | 0.02 |
| Biology | 30 | 0.16 | 0.10 | 0.09 | 0.04 | 0.10 | 0.11 | 0.20 | 0.12 | 0.00 | 0.00 | 0.05 | 0.03 |
| Neglected Tropical Diseases | 21 | 0.24 | 0.08 | 0.15 | 0.05 | 0.08 | 0.01 | 0.26 | 0.08 | 0.00 | 0.00 | 0.04 | 0.02 |

We checked whether there were differences in the proportions of *added, removed* and *unchanged* references made to the journal where the published version of the pair appeared. Figure S2 presents these results. Overall, the median references citing the publishing journal in all FOS fields was rather low (dots inside violins), while revisions during peer review reduced at least a share of references to the publishing journal (red violins). However, in engineering and technology, medical and health sciences, natural sciences and to a lesser extent agricultural sciences, most of the references to the published journal did not change (blue violins).

In all fields, there was a share of references added (green violins) to the publishing journal. On the one hand, this may show a pre-emptive tendency among authors to cite the published journal before submission, which then does not change during the peer review. On the other hand, in case of fields with highly specialised research themes, this might signal a narrow area of research that authors must cite to build their arguments based on few prior studies. We also checked the mean and median of the citations that *added, removed* or *unchanged* references accrued in the first three years after publication. While the median of the cited references was close to 10, the mean was closer to 100. We did not identify significant trends of, for instance, disproportionately adding highly cited references, which would reveal tactical signalling (e.g., citing highly cited references to corroborate findings).

This stable trend of authors citing the journal in which the published version appeared that remained unchanged during peer review could signal pre-emptive behaviour of authors to cite prior works published in the journals they aim to publish in, or may reflect that the journal is a key source or perhaps one of only a few journals in a narrow, specialised research area.

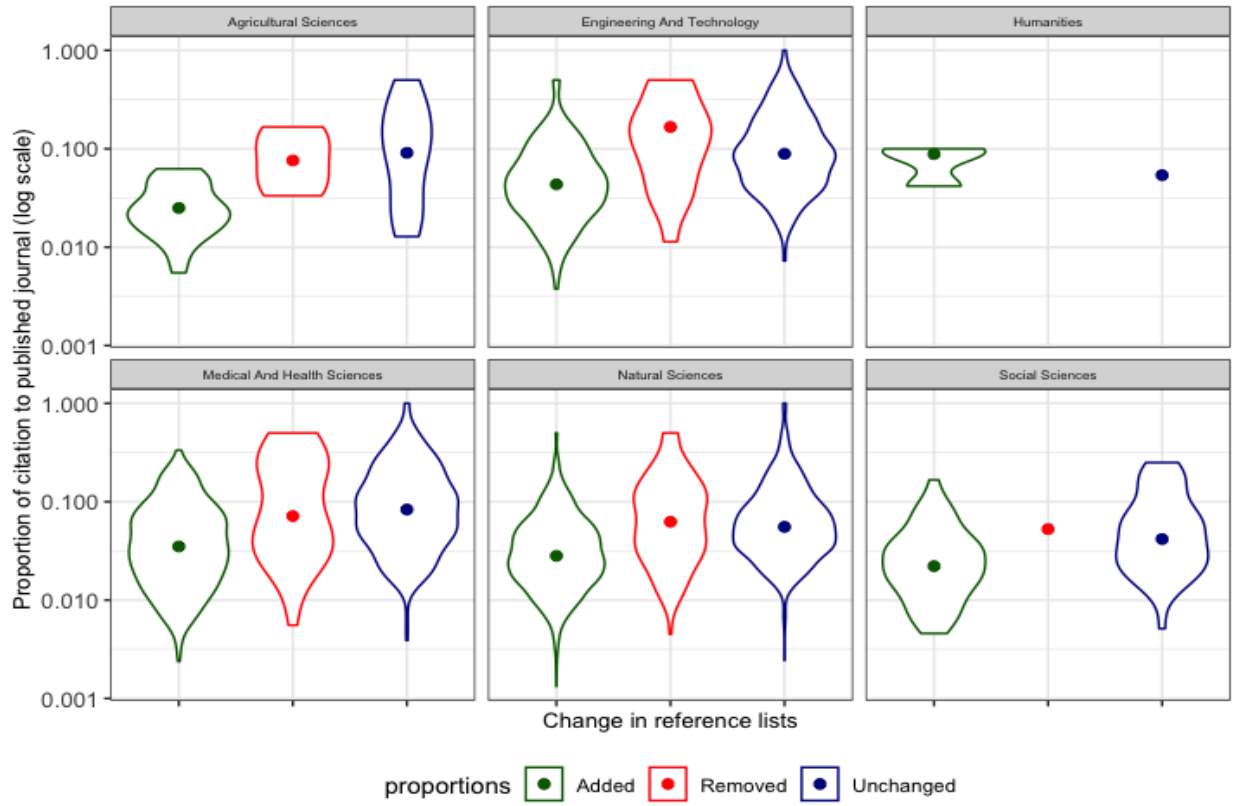

*Figure S2: Proportion of references to the publishing journal* (dots inside violins indicate median)

Figure S3 shows the distribution of *added* and *removed* references across document sections. *Added* references relate to the section of the publication and *removed* references relate to the preprint section. "Other section" includes non-standardised sections, such as acknowledgements. We included references in each section in which they were cited, as such references may be counted more than once and the denominator used for the percentage is the total number of references using multiple counting. Over one-third (35.5%, 82) of *removed* references and 1.7% (9) of *added* references could not be allocated to a section as, although they were in the reference list of the preprint or publication respectively, they were not actually cited in the text of the document. This issue occurred in 17 preprints, but was condensed primarily in 4 that had between 9 and 18 instances each.

These results validate the pattern seen in the main study that the majority of references were added to the introduction (30.0%) and discussion and conclusions sections (33.3%), with fewer changes in the methods (20.9%) and results sections (13.3%). The *removed* references here also follow this pattern, with the largest percentages removed from the introduction (29.4%) and discussion and conclusions (17.7%) over the methods (9.5%) and results (6.1%). Excluding the 82 references only removed as they were inaccurately included in the preprint reference list so that we may examine the distribution of "true" removals, nearly half of references were removed from the introduction (45.6%), 27.5% from discussion and conclusions, 14.8% from methods, and 9.4% from results.

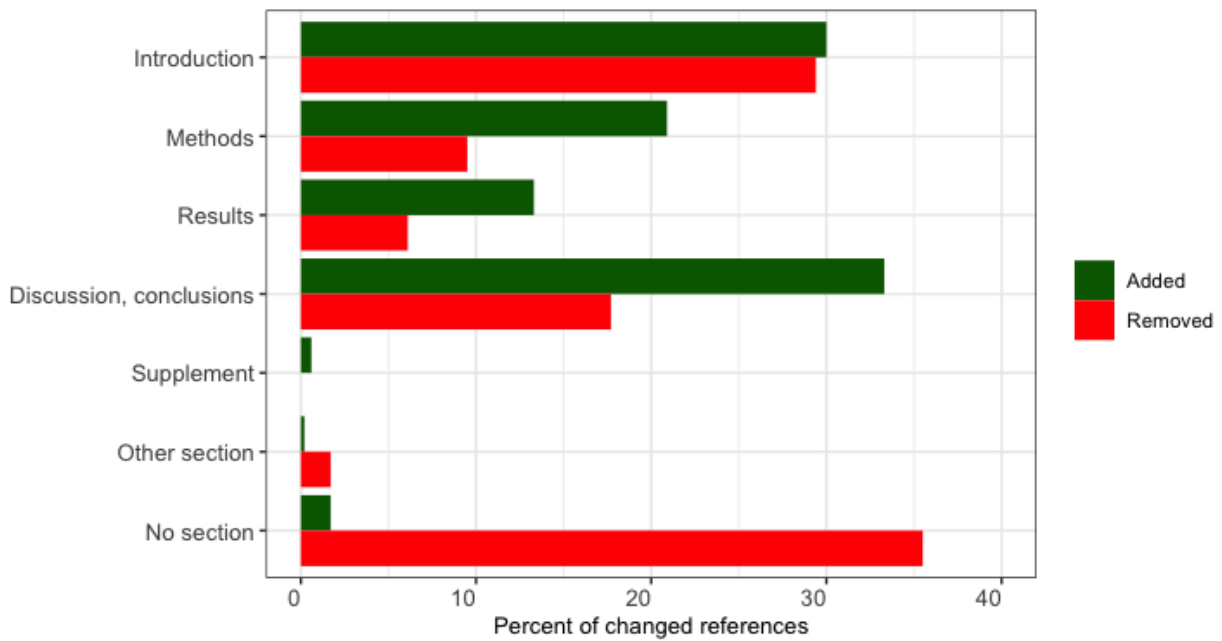

*Figure S3: Distribution of added and changed references in the validation sample by document section.*

Figure S4 shows the percentage of references added and removed by theme and section of the manuscript. Percentages are of the total number of added or removed references per manuscript section.

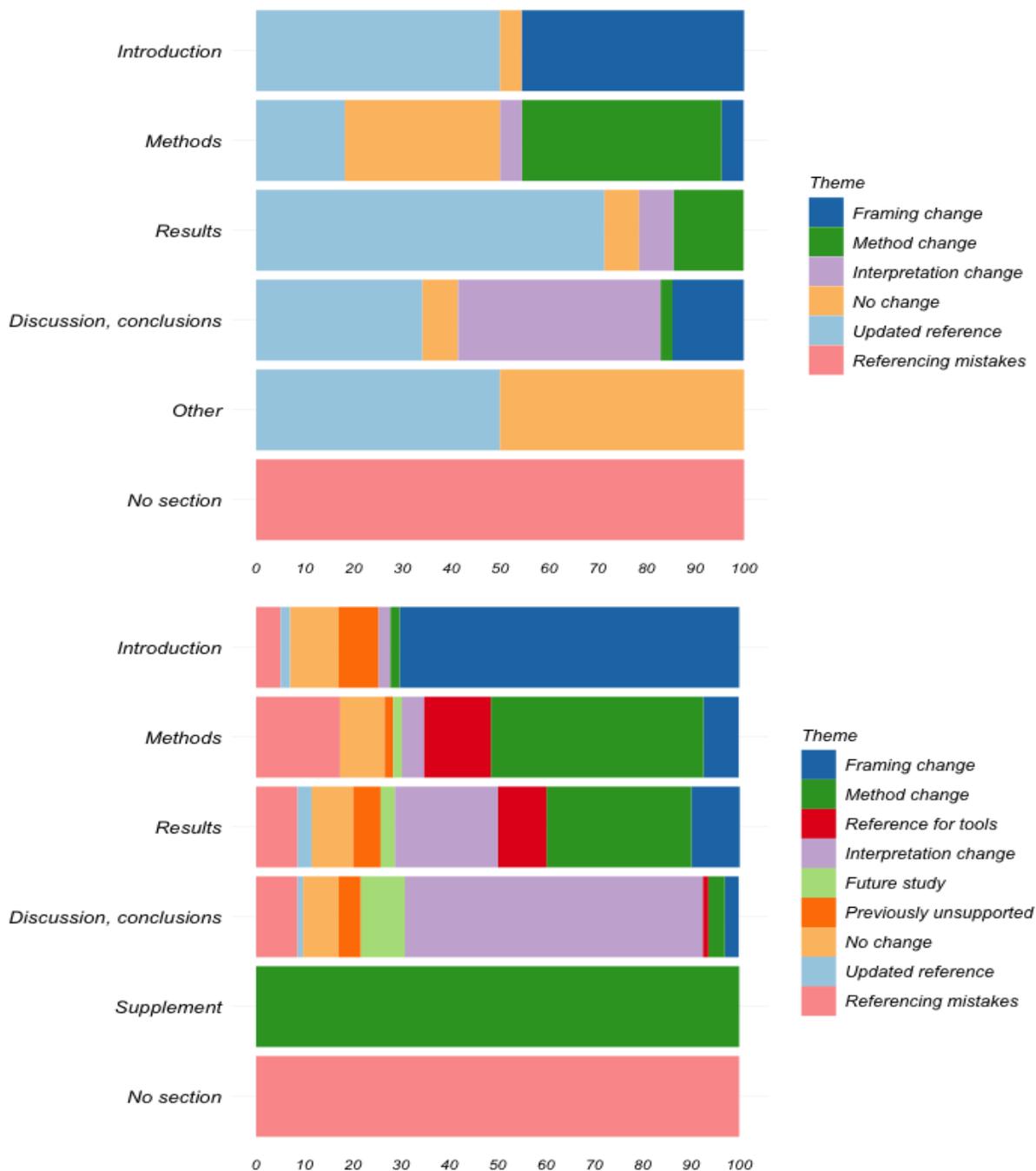

*Figure S4: The percentage of removed (top) or added (bottom) references per section by each thematic group.*